\documentclass[preprint,12pt]{elsarticle}

\usepackage{epsfig}

\usepackage{amssymb}

\journal{Journal}

\begin{document}

\begin{frontmatter}



\title{Two channel heat conduction in the superconducting state of the as-cast V$_{1-x}$Zr$_x$ alloys}


\author[1,2]{Sabyasachi Paul}
\author[1]{L. S. Sharath Chandra \corref{cor1}}
\cortext[cor1]{+91 731 2488522, email: lsschandra@rrcat.gov.in}
\author[1,2]{M. K. Chattopadhyay}
\address[1]{Free Eelectron Laser Utilization Laboratory, Raja Ramanna Centre for Advanced Technology, Indore, Madhya Pradesh - 452 013, India}
\address[2]{Homi Bhabha National Institute, Training School Complex, Anushakti Nagar, Mumbai 400 094, India}

\begin{abstract}
We present here the temperature dependence of heat capacity ($C$($T$)) and thermal conductivity ($\kappa$($T$)) in the superconducting state as well as in the  normal state of as-cast V$_{1-x}$Zr$_x$ alloys. Distinct jumps in the $C$($T$) of the alloys indicate the presence of three superconducting phases with transition temperatures $T_{C1}$ = 5.4~K, $T_{C2}$ = 8.2~K and $T_{C3}$ = 8.5~K. From the metallography micrographs, these three phases are identified to be  $\beta$-V, $\gamma$-ZrV$_2$, and $\gamma'$-ZrV$_2$ respectively. Apart from these phases, $\alpha$-Zr and $\beta$-Zr phases are also detected in these samples. The experimental $\kappa$($T$) in the superconducting state of these alloys is observed to be significantly higher than that expected theoretically. Our analysis suggests that the above observation is due to the coexistence of multiple superconducting and non superconducting phases which resulted in the two-parallel channels for the conduction of heat. 
 \end{abstract}


\end{frontmatter}

\section{Introduction}%

The C15 Laves phase (Hf,Zr)V$_2$ based superconductors have been considered to be promising materials for high field applications as an alternative to the Nb based alloys and compounds \cite{ino85, ten81, tac79} as well as for superconducting applications in the neutron radiation environment \cite{bro77, nas84}. Certain V-(Hf,Zr) composite tapes have exhibited a critical current density ($J_C$) larger than that of Nb$_3$Sn in fields higher than 14~T and at temperatures below 4.2~K \cite{tac79}. However, these C15 Laves phase materials are soft \cite{ino81} and brittle at room temperature and below \cite{ten81}, and need to be processed through engineering techniques for possible technological applications. Current interest in these materials stems from the recent developments in the processing technologies such as the powder in tube (PIT) technique, rapid heating and quenching (RHQ) and the rod restack process etc., for fabricating superconducting wires \cite{his04, seg16}. Recently, we have shown that the  multicomponent  V$_{1-x}$Zr$_x$ alloy superconductors are potential candidates for high field applications \cite{sha19}.
   
In this article we show that the electrons are the major carriers of heat in these as-cast V$_{1-x}$Zr$_x$ alloys. The thermal conductivity in the superconducting state of these alloys is larger than that expected for the Bardeen-Cooper-Schrieffer (BCS) type superconductors. Our analysis suggests that this discrepancy is due to the existence of two parallel channels for the conduction of heat in these alloys. 

\section {Experimental Details}%

The V$_{1-x}$Zr$_x$ ($x$ = 0-0.4) alloys were prepared by taking elemental vanadium (Alfa Aesar, 99.8\%) and zirconium (Alfa Aesar, 99.99\%) in stoichiometric proportion and melted together in a tri-arc furnace in Ar (99.999\%) atmosphere. The buttons were remelted six times to ensure homogeneity. The details of metallography can be found elsewhere \cite{sha19}. The temperature dependence of resistivity, heat capacity and thermal conductivity were measured in a 9~T Physical Property Measurement System (PPMS, Quantum Design, USA). 

\section {Results and Discussion}%

\begin{figure}[htb]
	\begin{center}
		\resizebox*{7cm}{!}{\includegraphics{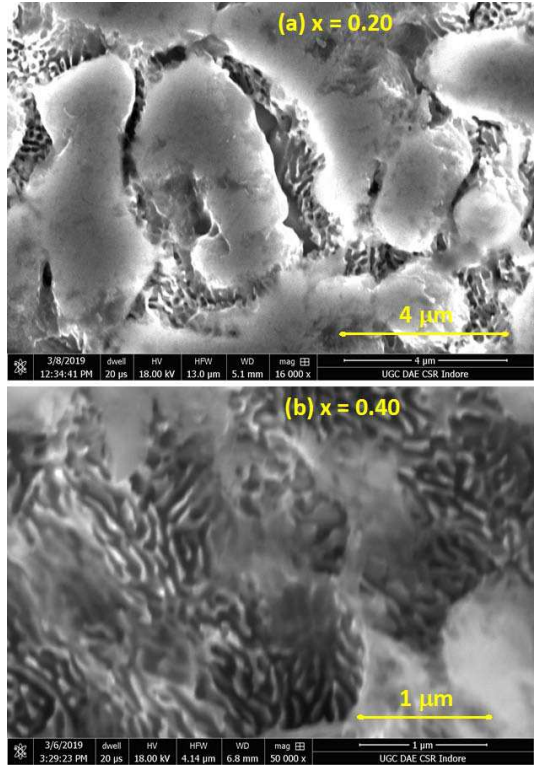}}%
		\caption {(colour online)  SEM metallographic images of the V$_{1-x}$Zr$_x$ alloys for $x$ = (a) 0.20 and (b) 0.40. During the preparation of the sample, temperature drops sharply from the molten state which leads to eutectic reaction even in these alloys. }
		\label{fig1}
	\end{center}
\end{figure}

The equilibrium phase diagram of the V$_{1-x}$Zr$_x$ alloys shows a peritectic isotherm at 1300~$^0$C for the 0.02$\leq x \leq$ 0.33 alloys  and an eutectic isotherm at 1230~$^0$C for the 0.33$\leq x \leq$1 alloys \cite{wil54}. In the peritectic reaction, solid $\beta$-V phase formed out of the melt reacts with the remaining liquid to form the $\gamma'$-ZrV$_2$ phase. In the eutectic reaction, the liquid being cooled transforms into $\beta$-Zr and $\gamma$-ZrV$_2$ phases \cite{ser05}. Upon further cooling, the $\beta$-Zr phase transforms below 777~$^0$C into the $\alpha$-Zr phase. The solubility of zirconium in vanadium at room temperature is only about 2\% \cite{ser05}. However, when cooled rapidly, the partial completion of the peritectic reaction results in the eutectic type reaction even in the 0.02$\leq x \leq$ 0.33 alloys\cite{sha19}. Therefore, five different structural phases namely $\beta$-V, $\gamma$-ZrV$_2$, $\gamma'$-ZrV$_2$, $\beta$-Zr and $\alpha$-Zr are observed in the present alloys \cite{sha19}. Figure 1 shows the high resolution scanning electron microscopy (SEM) images of $x$ = (a) 0.2 and (b) 0.4 alloys. The big plain grains in Fig. 1(a) and Fig. 1(b) are the $\beta$-V and $\gamma'$-ZrV$_2$ phases respectively. The lamellar structure in both these alloys contains $\gamma$-ZrV$_2$ and $\alpha$-Zr phases. The details of microstructures of the as-cast V$_{1-x}$Zr$_x$ alloys (non-equilibrium phase diagram) can be found elsewhere \cite{sha19}. 

Figure 2 shows the temperature dependence of heat capacity ($C$($T$))  of the V$_{1-x}$Zr$_x$ alloys in the absence of applied magnetic field. The $C$($T$) data for different alloys have been shifted vertically for clarity. We have observed that all the alloys with $x \geq$ 0.1 show three distinct features (marked by arrows) in the $C$($T$) below 10~K. A slope change and a small hump is observed at about 8.5~K whereas a clear jump in $C$($T$) is observed at about 8~K. The $\gamma$-ZrV$_2$ has a superconducting transition temperature in the range 8-9~K. The smallness of the change in $C$($T$) at about 8.5~K indicates that this feature is due to  the superconductivity in $\gamma'$-ZrV$_2$ phase. This is because of the fact that the volume fraction of the $\gamma'$-ZrV$_2$ phase formed around the grain boundaries of the $\beta$-V phase is quite small. The feature at about 8~K, which is prominent in the $x$ = 0.29, 0.33 and 0.40 alloys, is due to the superconductivity in the major phase $\gamma$-ZrV$_2$. The jump in the heat capacity at about 5.4~K is due to the superconductivity in the $\beta$-V phase. 
 
 \begin{figure}[htb]
 \begin{center}
 \resizebox*{7cm}{!}{\includegraphics{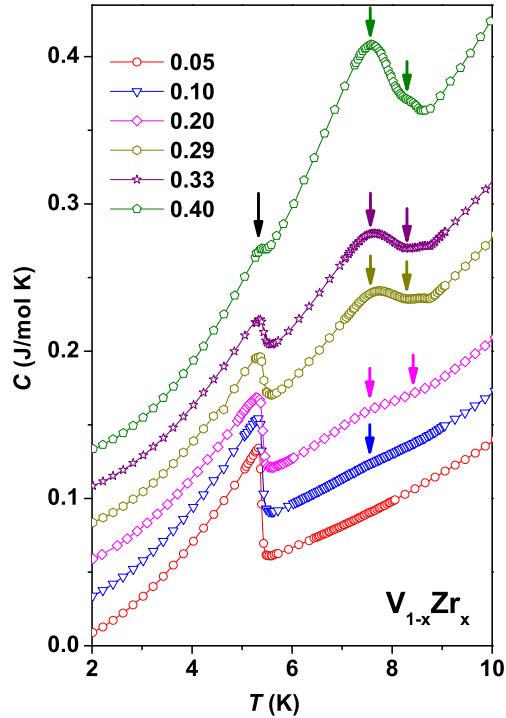}}%
 \caption {(colour online) Temperature dependence of heat capacity of the V$_{1-x}$Zr$_x$ alloys in the absence of  applied magnetic field. The data for different alloys have been shifted vertically. Three distinct features marked by arrows indicate the superconducting transitions. }
 \label{fig2}
 \end{center}
 \end{figure}
 
   \begin{figure}[htb]
   \begin{center}
   \resizebox*{7cm}{!}{\includegraphics{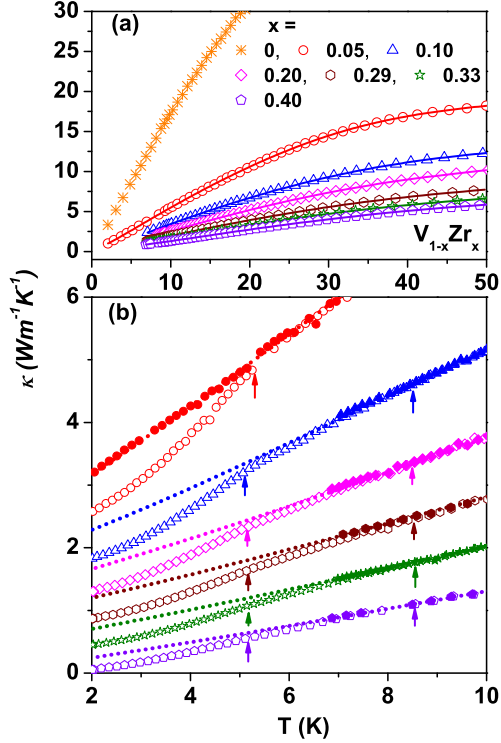}}%
   \caption {(colour online) (a) Temperature dependence of normal state $\kappa$($T$) of the V$_{1-x}$Zr$_x$  alloys in the temperature range 7-50~K. The open symbols are the experimental data. The solid lines represent the fit to the data. The $\kappa$($T$) reduces with increasing Zr-content. (b) The $\kappa$($T$) of the V$_{1-x}$Zr$_x$ alloys below 10~K measured in the absence of magnetic field (open symbols) and in the presence of 8~T  field (closed symbols). The dotted lines are the normal state $\kappa$ extrapolated in the range $T_C$ to 2~K from the fitting. The data for different alloys have been shifted vertically. The superconducting transition temperatures are marked by arrows.}
   \label{fig3}
   \end{center}
   \end{figure}
 
 Figure 3(a) shows the temperature dependence of the normal state $\kappa$($T$) of the V$_{1-x}$Zr$_x$ alloys. The open symbols are the experimental data. The $\kappa$($T$) decreases as the amount of zirconium in vanadium is increased. For all the alloys, $\kappa$($T$) is linear below 20~K. Figure 3(b) shows the temperature dependence of thermal conductivity ($\kappa$($T$)) of the V$_{1-x}$Zr$_x$ alloys below 10~K measured in the absence of magnetic field (open symbols) as well as in the presence of 8~T  field (closed symbols). The data for different alloys have been shifted vertically for clarity. In the absence of applied magnetic field, the $\kappa$($T$) starts to decrease when the temperature is decreased below $T_{C3}$ = 8.5~K. These results indicate that the electrons are the major carrier of heat in these alloys. Since the change in $\kappa$($T$) across the $T_C$ is quite small, distinct features at about 8~K and 8.5~K are not visible. 
 
  \begin{figure}[htb]
  \begin{center}
  \resizebox*{7cm}{!}{\includegraphics{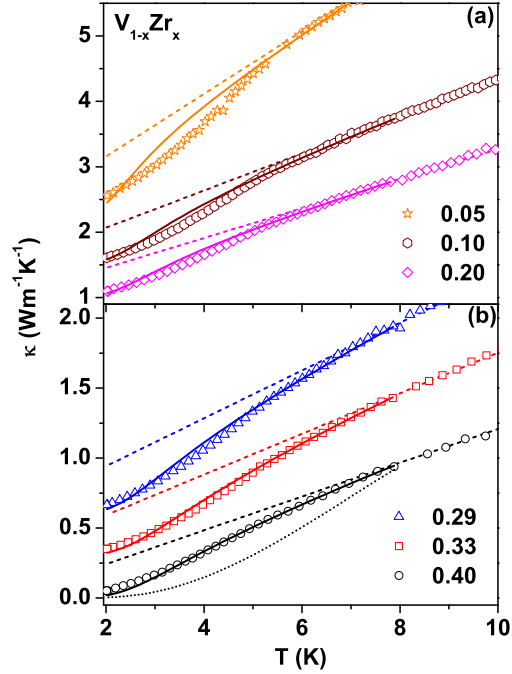}}%
  \caption {(colour online)  Results of the analysis of the $\kappa_{es}$($T$) of the V$_{1-x}$Zr$_x$ alloys. The open symbols represent experimental data. The dashed line corresponds to the normal state data. The dotted line in (b) is the $\kappa_{ei,s}$ obtained from BCS theory for the $x$ = 0.40 alloy. The $\kappa_{ei,s}$ is quite small in comparison to the experimental data. This is true for other alloys as well. The solid lines are the fit to the data using a parallel channel model for heat conduction. }
  \label{fig4}
  \end{center}
  \end{figure}
 
  For a superconductor, the $\kappa$($T$) can be expressed as \cite{tri04}, 
 \begin{equation}
 \kappa_z = \kappa_{ez}+\kappa_{lz},
 \end{equation}
 where the electronic part of the thermal conductivity $\kappa_{ez}$ limited by the scattering from impurities and phonons can be expressed as \cite{tri04}  
 \begin{equation}
 \kappa_{ez}^{-1} = \kappa_{ei,z}^{-1}+\kappa_{el,z}^{-1}.
 \end{equation}
 Here, $\kappa_{ei,z}^{-1}$ and $\kappa_{el,z}^{-1}$ are the thermal resistivities due to the scattering of electrons from the defects and phonons respectively. The subscript $z$ = $n$ represents the normal state and $z$ = $s$ represents the superconducting state. In the normal state, $\kappa_{ei,n}^{-1} = AT^{-1}$ and $\kappa_{el,n}^{-1} = BT^2+\mathcal{O}(T^ 4)$ where $A$ and $B$ are constants and the exact form of  $\kappa_{el,n}$ can be found in Ref. \cite{tri04}.
 The phonon thermal conductivity $\kappa_{lz}$ is given by \cite{sha12, sab19}
 \begin{equation}
 \kappa_{lz} =  MT^3 \int^{\infty}_{0} dx \frac{ x^4 e^x (e^x -1)^{-2}}{N + L x T + Cg(x) x T + P x^4 T^4} ,
 \end{equation}
 where $M = k_B^4 /(2\pi^2 \hbar^3 v_s)$, $k_B$ is the Boltzmann constant, $v_s$ is the sound velocity and $\hbar$ is the reduced Plank's constant. The coefficients $N$, $L$, $C$ and $P$ represent the strength of phonon scattering due to boundaries, dislocations, electrons and point defects respectively. The $g(x)$ is the ratio of the phonon-electron scattering cross-sections in the superconducting and normal states \cite{bar59} and $x$ = $\hbar/k_BT$ is the phonon frequency. The $g(x)$ goes to unity in the normal state.
 
 The metallography images shown in the Fig.1 as well as the analysis of the pinning force density of these alloys \cite{sha19} indicate that the grain boundaries and the point defects influence the properties of these alloys. Therefore, we have taken $C$ = 0 and $L$ = 0 for the analysis of the $\kappa_n$($T$). Since, the Debye temperature ($\theta_D$) cannot be extracted reliably from the temperature dependence of heat capacity due to the presence of multiple structural phases, we have used $\theta_D$ as a parameter in fitting the $\kappa_n$($T$).  Thus, the fitting of the $\kappa_n$ is carried out in the temperature range 6-50~K using five parameters: $A$, $B$, $\theta_D$, $N$ and $P$. 
 
The solid lines Fig.3(a) represent the fit to the $\kappa_n$ and the parameters are listed in Table 1. We found that the data can be fitted best with these parameters only. The results of the analysis of the $\kappa$($T$) of vanadium is presented elsewhere \cite{sab19}. We see that the electrons are the major carriers of heat in the normal state of V$_{1-x}$Zr$_x$ alloys. The static defects such as grain boundaries and point defects limit the heat carried by the electrons. We also found that the phonons are ineffective in limiting the electronic heat flow below 10~K as $A >> B$. The dotted lines in Fig.3(b)  are the extrapolated normal state data (since the alloys do not become normal below 6~K in 8~T field) at low temperatures.  The heat conduction by phonons at a temperature below $T_C$ is same (due to $C$ = 0) in both the superconducting and normal states of these alloys. Therefore, the variation of $\kappa_s$(T) is dictated by $\kappa_{ei,s}$

 \begin{table*} 
\centering
\caption{\label{tab:table1} Values of parameters obtained by fitting $\kappa_n$($T$) and $\kappa_s$($T$).}
\vskip 10 mm
\begin{tabular}{cccccccc}
\hline
x& $A$  & $B$ & $\theta_D$ & $N/M$&$P/M$&$c$ \\
&(mW$^{-1}$K$^{2}$)&(mW$^{-1}$K$^{-1}$)&(K)&(mW$^{-1}$K$^{4}$)&(mW$^{-1}$)&\\
\hline
0.05&2.09&5.99$\times$10$^{-6}$&421&8.53$\times$10$^3$&0.017&0.95 \\
0.10& 3.52&7.24$\times$10$^{-6}$&430&1.15$\times$10$^4$&0.017&0.9  \\
0.20& 4.39&8.14$\times$10$^{-6}$&397&4.54$\times$10$^4$&0.013&0.9  \\
0.29&5.84&9.19$\times$10$^{-6}$&415&3.99$\times$10$^4$&0.024&0.85  \\
0.33&6.87&11.32$\times$10$^{-6}$&397&9.07$\times$10$^4$&0.018&0.8  \\
0.40&8.27&10.66$\times$10$^{-6}$&428&1.88$\times$10$^5$&0.011&0.8  \\
\hline
\end{tabular}
\end{table*}

To analyse the $\kappa_s$($T$) in the superconducting state, we have subtracted the $\kappa_{ln}$ from the $\kappa_s$ to obtain $\kappa_{es}$($T$). The $\kappa_{ei,s}$ is given by Bardeen-Rickayzen-Tewordt \cite{bar59} as

\begin{eqnarray}\nonumber
\frac{\kappa_{ei,s}}{ \kappa_{ei,n} } &=&R_{ei} \\
&=& \frac{1}{f(0)} [f(-y) + y ln(1 + exp(-y)) +\frac{y^2}{2(1+exp(y))}]
\end{eqnarray}
where $y=\frac{\Delta(T)}{k_B T}$ ($\Delta(T)$ is the superconducting energy gap), and the Fermi integral $f(-y)$ is given by $f(-y) = \int_{0}^{\infty} \frac{zdz}{1+exp(z+y)}$.

 Figure 4 shows the analysis of the $\kappa_{es}$($T$) of the V$_{1-x}$Zr$_x$ alloys. The open symbols represent the experimental data. The dashed line is the $\kappa_{en}$. All the samples show deviation from the normal state data below 8.5~K which corresponds to the $T_C$ of the $\gamma'$-ZrV$_2$ phase. Since the change in $\kappa$ just below $T_C$ is quite small, we can assume that the $T_C$ of both $\gamma'$-ZrV$_2$ and $\gamma$-ZrV$_2$ are same and equal to 8.5~K for the analysis of $\kappa_{es}$. The deviation from the normal state data below 8.5~K increases with increasing $x$. The dotted line in (b) is the $\kappa_{ei,s}$ for the $x$ = 0.40 alloy obtained using eq.4. To estimate $\kappa_{ei,s}$, we have used $\Delta/k_BT_C$ = 1.9 obtained from the analysis of $C$($T$) of an annealed ZrV$_2$ sample which has only $\gamma$ and $\gamma'$ phases (not shown here). We find that experimental value of $\kappa_{es}$($T$) of $x$ = 0.40 alloy is higher than the $\kappa_{ei,s}$ obtained from the BCS theory and is true for other alloys as well. This suggest that there is an additional channel for heat conduction in these samples. The phases such as $\alpha$-Zr, $\beta$-V and $\beta$-Zr are not superconducting in the temperature range about 5-8.5~K. Therefore, in this temperature range, a normal channel for heat conduction is available in parallel to the superconducting channel. In such cases, by taking two conducting channels in parallel, the temperature dependence of $\kappa_{ei,s}$ can be expressed as 
 
 \begin{equation}
 \kappa_{es}^{-1} = (1-c)\kappa_{ei,s}^{-1} +c \kappa_{ei,n}^{-1}.  
 \end{equation}

 Therefore,
 
  \begin{equation}
  \kappa_{es} = \frac{\kappa_{ei,n}R_{ei}}{cR_{ei}+(1-c)},
  \end{equation}
 
 where $c$ is the weight factor for the thermal resistivity in the normal state of ZrV$_2$ phase. The solid lines are the fit to the data obtained using eq.5. The values  of $c$ are given in table 1. It is observed that $c$ = 0.95 for $x$ = 0.05 and it decreases with the increasing $x$. For $x$ = 0.40 alloy, the $c$ turns out to be 0.8. The $\kappa_{es} \approx \kappa_{ei,n}$  at $T \sim T_C = T_{C3}$ as $cR_{ei} >>(1-c)$ ($R_{ei} \approx 1$ at $T = T_C$). On the other hand, the $\kappa_{es}  \approx \kappa_{ei,s}$ at $T <<T_C = T_{C3}$ as $cR_{ei} << (1-c)$ ($R_{ei} \approx 0$ at $T = T_C$). Figure 4(a) shows that there is a significant deviation of the fitted curve from the experimental data below 5~K for $x$ = 0.05 and 0.10 alloys. In these alloys, the volume fraction of the $\beta$-V phase is quite large \cite{sha19} and hence, the thermal transport below 5~K is dominated by the $\beta$-V phase. The volume fraction of the $\beta$-V phase is quite small for $x$ = 0.20, 0.29 and 0.33 and it is insignificant for $x$ = 0.40 \cite{sha19}. Therefore, the fit using the parallel channel model is good over the entire temperature range of measurement for the alloys with $x \geq$ 0.20. These findings are significant towards the realization of the V-Zr based superconducting wires for high field applications. 
 
 \section {Conclusions}%

In conclusion, the experimental thermal conductivity in the superconducting V$_{1-x}$Zr$_x$ alloys is observed to be larger than that expected from the BCS theory.  We show that the coexistence of multiple superconducting and non superconducting phases results in the two-parallel channels for heat conduction. We have also shown that in the superconducting state, more than 80\% of the heat is carried by the normal channel. These findings along with the large $J_C$ make these alloys as potential candidates for high magnetic field applications. 






\begin{thebibliography}{15}

\bibitem{ino85} K. Inoue, T. Kuroda, and K. Tachikawa, Superconducting properties of V$_2$(Hf,Zr) laves phase multifilamentary wires, J. Nucl. Mat. 133\&134 (1985) 815 -818.
\bibitem{ten81} M. Tenhover, High critical currents obtained by heat treating rapidly quenched Hf-Zr-V metallic glasses, IEEE Trans. Mag. 17 (1981) 1021-1024.
\bibitem{tac79} K. Tachikawa, Research and development at the Japanese national research institute for metals, Cryogenics, 19 (1979) 307-315.
\bibitem{bro77} B. S. Brown, J. W. Hafstrom and T. E. Klippert, Changes in the superconducting critical temperature after fast-neutron irradiation, J. Appl. Phys. 48 (1977) 1759-1761.
\bibitem{nas84} I. A. Naskidashvili, L. S. Topchyan, A. I. Naskidashvi-Li and E. S. Makarenkov, Influence of irradiation on the superconducting transition temperature of intermetallic V$_2$Zr, Rad. Effects Lett. 85 (1984) 97-102.
\bibitem{ino81} K. Inoue, H. Wada, T. Kuroda, and K. Tachikawa, Stress effects on superconducting properties of the composite-processed V$_2$(Hf,Zr), Appl. Phys. Lett. 38 (1981) 939-941.
\bibitem{his04} Y. Hishinuma, A. Kikuchi, Y. Iijima, Y. Yoshida, T. Takeuchi, A. Nishimura and K. Inoue, The fabrication of a V-based Laves phase compound superconductor through a rapid heating and quenching process, Supercond. Sci. Technol. 17 (2004) 1031-1036.
\bibitem{seg16} C. Segal, C. Tarantini, Z. H. Sung, P. J. Lee, B. Sailer, M. Thoener, K. Schlenga, A. Ballarino, L. Bottura, B. Bordini, C. Scheuerlein, and D. C. Larbalestier, Evaluation of critical current density and residual resistance ratio limits in powder in tube Nb$_3$Sn conductors, Supercond. Sci. Technol. 29 (2016) 085003-1-10.
\bibitem{sha19} L. S. Sharath Chandra, S. Paul, A Khandelwal, A. Sagdeo, R. Venkatesh, Kranti Kumar, A. Banerjee and M. K. Chottopadhyay, Evolution of high field superconductivity and high critical current density in the as-cast V$_{1-x}$Zr$_x$ alloys, arXiv:1908.07288 (2019).
\bibitem{wil54} J. T. Williams, Retrospective Theses and Dissertations. Paper 13305 ({\it The vanadium-zirconium alloy system}, Iowa state collage) pp.1-73 (1954).
\bibitem{ser05} C. Servant, Thermodynamic assessments of the phase diagrams of the Hanium-Vanadium and Vanadium-Zirconium systems, J. Phase Equilib. Diff. 26 (2005) 39-49.
\bibitem{tri04} See e. g. Tritt T M 2004 {\it Thermal Conductivity: Theory, Properties and Applications} (Kluwer Academic/Plenum Publishers, New York).
\bibitem{sha12} L. S. Sharath Chandra, M. K. Chattopadhyay, S. B. Roy, V. C. Sahni and G. R. Myneni, Magneto thermal conductivity of superconducting Nb with intermediate level of impurity, Supercond. Sci. Technol. 25 (2012) 035010-1-12.
\bibitem{sab19} S. Paul, L. S. Sharath Chandra and M. K. Chattopadhyay,  Renormalization of electron-phonon coupling in the Mott-Ioffe-Regel limit due to point defects in the V$_{1-x}$Ti$_x$ alloy superconductors, J. Phys.: Condens. Mater, 31 (2019) 475801-1-9. 
\bibitem{bar59} J. Bardeen, G. Rickayzen and L. Tewordt, Theory of the thermal conductivity of superconductors, Phys. Rev. 113 (1959) 982-994.



\end{thebibliography}







\label{lastpage}

\end{document}